# Simulation Analysis of Routing Protocols using Manhattan Grid Mobility Model in MANET

Youssef Saadi[1], Said El Kafhali[1, 2], Abdelkrim Haqiq[1, 2], Bouchaib Nassereddine[1]

[1] Computer, Networks, Mobility and Modeling laboratory
Department of Mathematics and Computer
FST, Hassan 1st University, Settat, Morocco

[2] e-NGN Research group, Africa and Middle East

## ABSTRACT
A Mobile Ad-hoc Network (MANET) is a self-configuring infrastructure less network of mobile devices connected by wireless links. In this network technology, simulative analysis is a significant method to understand the performance of routing protocols. In this paper three protocols AODV, DSDV and DSR were simulated using Manhattan Grid Mobility Model.

The reactive (AODV, DSR) and proactive (DSDV) protocol's internal mechanism leads to considerable performance difference. The performance differentials are analyzed using NS-2 which is the main network simulator, NAM (Network Animator), AWK (post processing script) and were compared in terms of Packet Delivery Fraction (PDF), Average end-to-end Delay and Throughput, in different environments specified by varying network load, mobility rate and number of nodes.

Our results presented in this research work demonstrate the performance analysis of AODV, DSDV and DSR routing protocols. It has been observed that, under Manhattan Grid mobility model, AODV and DSR performs better than DSDV in terms of PDF and Throughput. However in term of Average end-to-end Delay, DSDV appears to be the best one.

## 1. INTRODUCTION
A Mobile Ad hoc Network (MANET) is autonomous, self-configuring network of mobile nodes that can be set up randomly and formed without the need of any existing network infrastructure or centralized administration. All nodes can be mobile resulting in a possibly dynamic network topology which is a real challenging issue in mobile ad hoc networks.

The dynamic nature of MANET topology imposes the use of efficient routing protocols that ensure the delivery of packets safely to their destinations with acceptable delays.
Simulation studies of MANET routing protocols have mostly considered Random Waypoint as a reference mobility model [17, 18].

In order to examine many different MANET applications, there is a need to provide additional mobility models. There are various mobility models such as Random Way Point, Manhattan Grid Mobility Model, Reference Point Group Mobility Model (RPGMM), Freeway Mobility Model, Gauss Markov Mobility Model etc that have been suggested for evaluation [3, 6].

Many researches have been focused on the evaluation of routing protocols according to nodes mobility: a performance comparison of DSR and AODV protocols based on Manhattan Grid (MG) model has been published in [12]. A performance study of DSR and AODV considering probabilistic random walk and boundless simulation area has been presented in [13]. A performance evaluation of DSDV and AODV using scenario based mobility models has been presented in [2]. A comparative analysis of DSR and DSDV protocols, considering Random Waypoint, Group Mobility, Freeway and MG models can be found in [10], Performance Analysis and Comparison of MANET Routing Protocols vs. Mobility Models is presented in [20].

In our work, we have selected the Manhattan Grid mobility model that models a movement in city streets environment. MG model uses a grid topology that represents streets within a city so as to simulate movement in urban area. In this model, the nodes move in vertical or horizontal direction on an urban map.

Related to this scenario, we have investigated the performance of AODV, DSR (On-Demand routing protocol) and DSDV (proactive routing protocol) for performance comparison.
The purpose of this work is to understand their working mechanism and to show which routing protocol performs better under constraints of network size, mobility rate and network load.

The rest of this paper is organized as follows. Section 2 describes the AODV, DSR and DSVD routing protocols. The simulation environment and performance parameters are described in Section 3. In Section 4 we present simulation results and analysis. Finally, Section 5 concludes the paper.

## 2. ROUTING PROTOCOLS DESCRIPTION
Three routing protocols are considered in this paper, namely; AODV, DSR and DSDV. Below is a brief description of each protocol:

### 2.1 Ad Hoc on-Demand Distance Vector Routing (AODV)
The Ad Hoc on-Demand Distance Vector Routing (AODV) [8] is a routing protocol for mobile ad hoc

networks (MANETs) and other wireless ad-hoc networks provides on-demand route discovery. It is a reactive routing protocol, meaning that it establishes a route to a destination only on demand. Whenever the nodes need to send data to the destination, if the source node doesn't have routing information in its table, route discovery process begins to find the routes from source to destination. A node requests a route to a destination by broadcasting an RREQ message to all its neighbors. RREQ message comprises broadcast ID, two sequence numbers, the addresses of source and destination and hop count. The intermediary nodes which receive the RREQ message could do two steps: If it isn't the destination node then it'll rebroadcast the RREQ packet to its neighbors. Otherwise it'll be the destination node and then it will send a unicast replay message, route replay (RREP), directly to the source from which it was received the RREQ message. This RREP is unicast along the reverse-routes of the intermediate nodes until it reaches the original requesting node. This process repeats until the RREQ reaches a node that has a valid route to the destination.

At each node [19], AODV maintains a routing table. Each node has a sequence number. When a node wants to initiate route discovery process, it includes its sequence number and the most fresh sequence number it has for destination. The intermediate node that receive the RREQ packet, replay to the RREQ packet only when the sequence number of its path is larger than or identical to the sequence number comprised in the RREQ packet. A reverse path from the intermediate node to the source forms with storing the node's address from which initial copy of RREQ. Thus, at the end of this request-response cycle a bidirectional route is established between the requesting node and the destination. When a node loses connectivity to its next hop, the node invalidates its route by sending an RERR to all nodes that potentially received its RREP.

As long as the route remains active, it will continue to be maintained. A route is considered active as long as there are data packets periodically travelling from the source to the destination along that path. Once the source stops sending data packets, the links will time out and eventually be deleted from the intermediate node routing tables. When a source node wants to send data to some destination, first it searches the routing table; if it can find it, it will use it. Otherwise, it must start a route discovery to find a route [1]. It is also Route Error (RERR) message that used to notify the other nodes about some failures in other nodes or links [15].

## 2.2 Dynamic Source Routing (DSR)
The Dynamic Source Routing (DSR) [9] is a reactive routing protocol designed specifically for use in multi-hop wireless ad hoc networks of mobile nodes. In this protocol each source determines the route to be used in transmitting its packets to selected destinations. There are two main components, called Route Discovery and Route Maintenance. Route Discovery is the mechanism by which a node wishing to send a packet to a destination obtains a path to the destination. Route Maintenance is the mechanism by which a node detects a break in its source route and obtains a corrected route. The sender knows the complete hop by hop route to the destination. These routes are stored in a route cache [5, 14]. The protocol allows multiple routes to any destination and allows each sender to select and control the routes used in routing its packets, for example for use in load balancing or for increased robustness. The DSR protocol is designed mainly for mobile ad hoc networks of up to about two hundred nodes, and is designed to work well with even very high rates of mobility.

## 2.3 Destination Sequenced Distance Vector (DSDV) Protocol
The Destination Sequenced Distance Vector routing protocol [7] is a proactive routing protocol based on the Bellman-Ford routing algorithm. It was developed by C. Perkins and P.Bhagwat in 1994 [16]. This protocol adds a new attribute, sequence number, to each route table entry at each node. Each node in the mobile network maintains a routing table in which all of the possible destinations within the non-partitioned network and the number of routing hops to each destination are recorded. In this protocol, packets are routed between nodes of an ad hoc network using routing tables stored at each node. Each routing table, at each node, contains a list of the addresses of every other node in the network. Along with each node's address, the table contains the address of the next hop for a packet to take in order to reach the node. This protocol was motivated for the use of data exchange along changing and arbitrary paths of interconnection which may not be close to any base station.

## 3. SIMULATION ENVIRONMENT
### 3.1 Simulation Model
The network simulations have been carried out using Network Simulator version 2 (NS-2.34) and its associated tools for animation and analysis of results.

We chose a Linux platform i.e. UBUNTU 10.10, as Linux offers a number of programming development tools that can be used with the simulation process.

We analyzed the experimental results contained in generated output trace files by using the AWK command.

We have generated mobility scenarios for Manhattan Grid Model using the BONNMOTION [4] tool and have converted generated scripts to the supported ns2 format so that they can be integrated into TCL scripts.

Random traffic connections of CBR and TCP can be setup between mobile nodes using a traffic-scenario generator script (cbrgen.tcl) [11]. It can be used to create CBR and TCP traffic connections between wireless mobile nodes. In order to create a traffic-connection file, we need to define the type of traffic connection (CBR), the number of nodes and maximum number of connections to be setup between them. CBR is generally used to simulate multimedia traffic on limited capacity channels, or to fill in background traffic to affect the performance of other applications being analyzed. The TCP sources are not being chosen because they adapt to the load of the network.

The simulations carried out, traffic models were generated for 10, 30 and 50 nodes with CBR traffic sources, with maximum connections of 8, 25 and 40 at a rate of 4 packets per second.

Mobility models were created for the simulations using 10, 30 and 50 nodes, with pause times of 0, 20, 40, 60, 80 and 100 seconds, minimum speed of 5m/s and maximum speed of 20m/s, topology boundary of 500x500 and simulation time of 100secs.

## 3.2 Simulation parameters
The simulation parameters are listed in Table 1.

**Table 1: Simulation parameters**

| Parameter | Value |
|---|---|
| Simulator | NS-2 (Version 2.34) |
| Channel type | Channel/Wireless channel |
| protocols | AODV, DSR and DSDV |
| Simulation duration | 100s |
| Number of nodes | 10, 30, 50 |
| Transmission range | 250m |
| Movement Model | Manhattan Grid Model |
| MAC Layer Protocol | 802.11 |
| Pause Time (s) | 0, 20, 40, 60, 80, 100 |
| Maximum speed | 20m/s |
| Minimum speed | 5m/s |
| Packet Rate | 4 packet/s |
| Traffic type | CBR |
| Data Payload | 512 bytes/packet |
| Max of CBR connections | 8, 25, 40 |

## 3.3 Performance Parameters
This paper analysed the following important performance parameters for compared the AODV, DSR and DSDV routing protocols:

### 3.3.1 Packet Delivery Fraction (PDF)
It is the ratio of all received data packets successfully at destinations and all data packets sent by CBR sources.

### 3.3.2 Average end-to-end Delay
It represents the delay encountered between the sending and receiving of the packets.

It is the time from the transmission of data packet at a source node until packet delivery to a destination which includes all possible delays caused by:

- Buffering during route discovery process
- Retransmissions delays
- Queuing at Interface Queue
- Propagation and transfer times of data packet.

### 3.3.3 Throughput
It is the average number of messages successfully delivered per unit time.

## 4. SIMULATION RESULTS AND ANALYSIS
## 4.1 Simulation Results

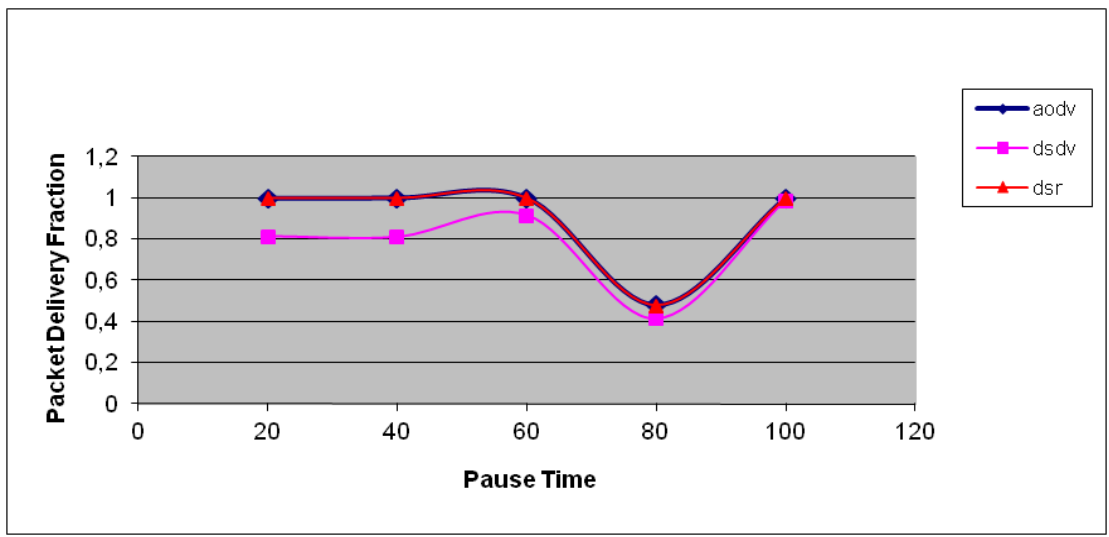

**Fig 1: PDF under Pause Time (fixed 10 nodes).**

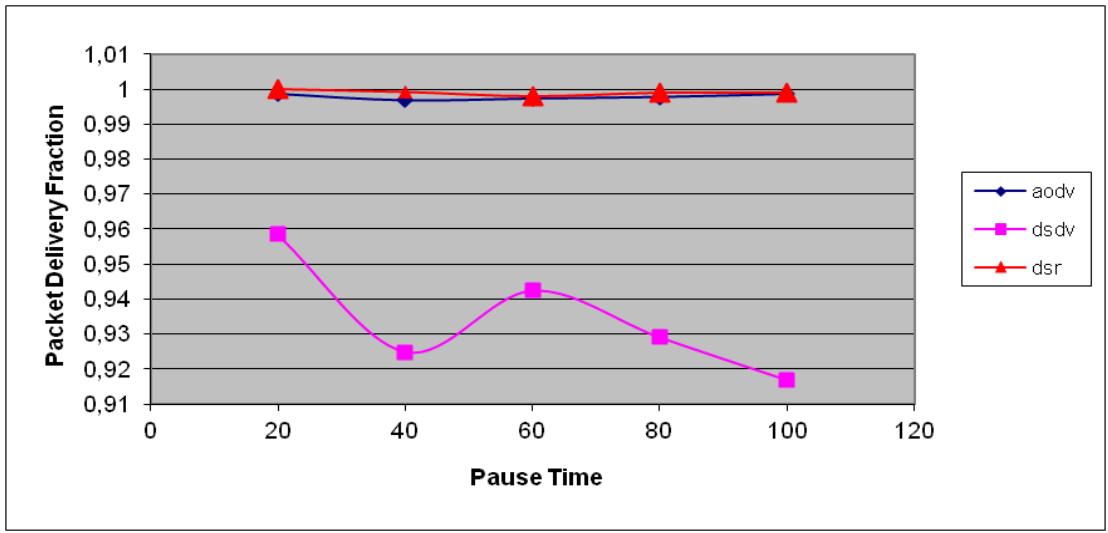

**Fig 2: PDF under Pause Time (fixed 30 nodes).**

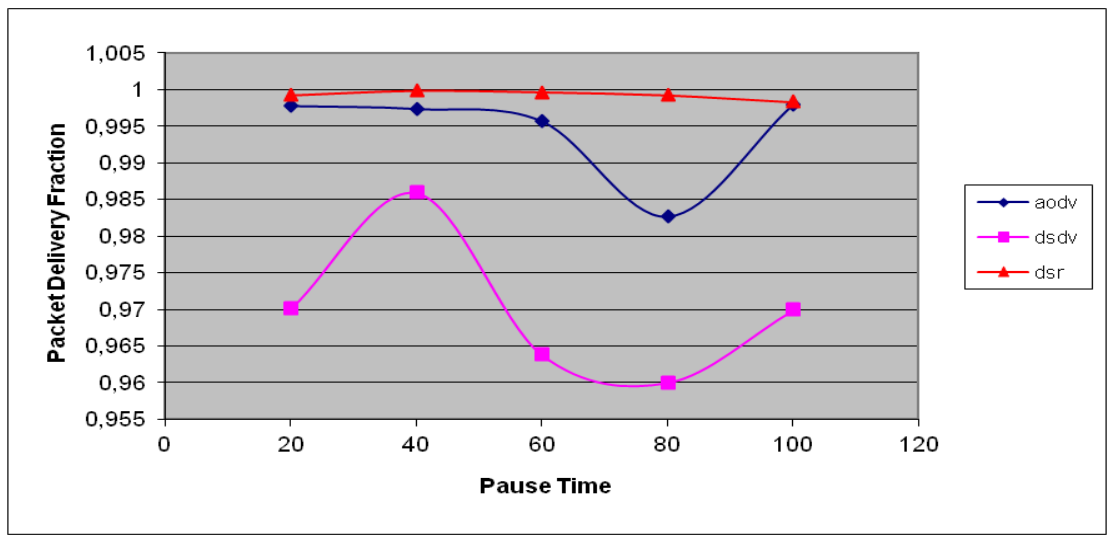

**Fig 3: PDF under Pause Time (fixed 50 nodes).**

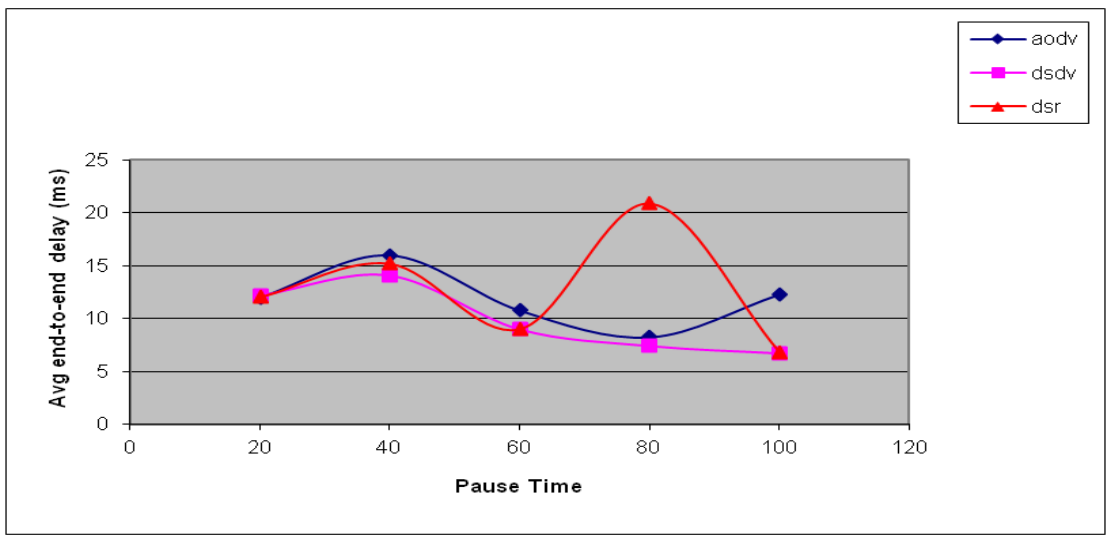

**Fig 4: Average end-to-end Delay under Pause Time (fixed 10 nodes).**

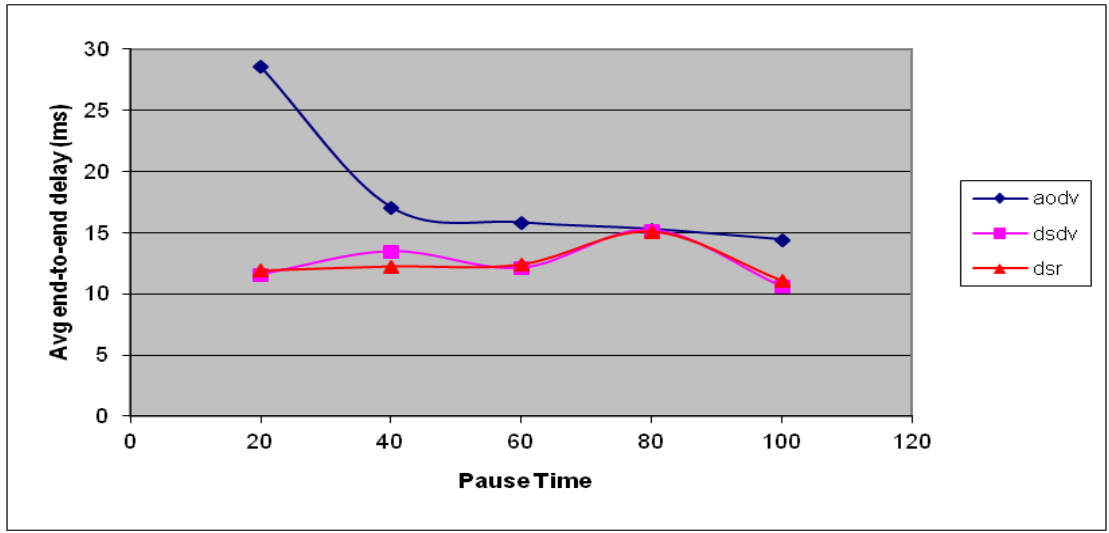

**Fig 5: Average end-to-end Delay under Pause Time (fixed 30 nodes).**

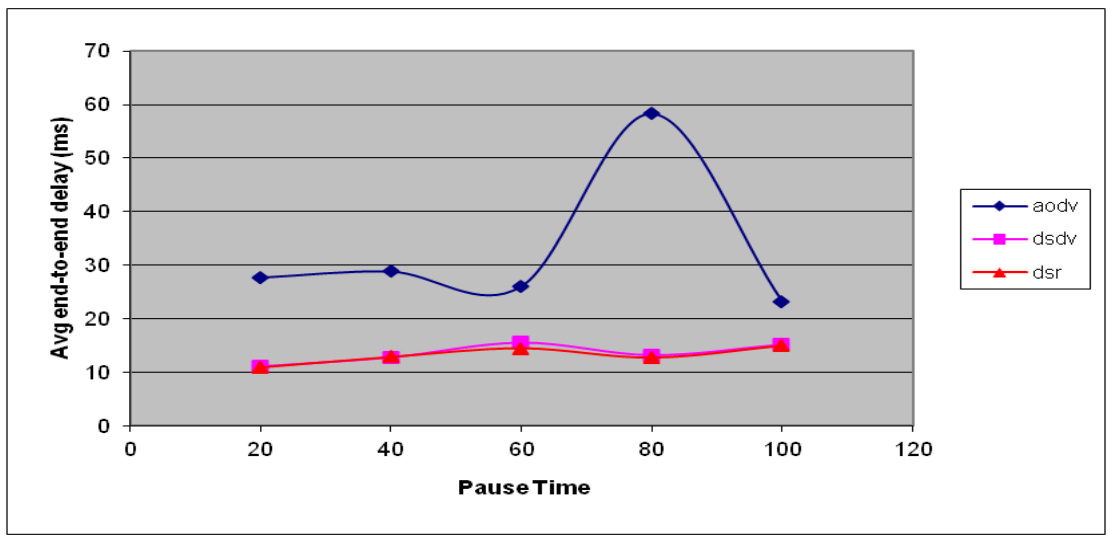

**Fig 6: Average end-to-end Delay under Pause Time (fixed 50 nodes).**

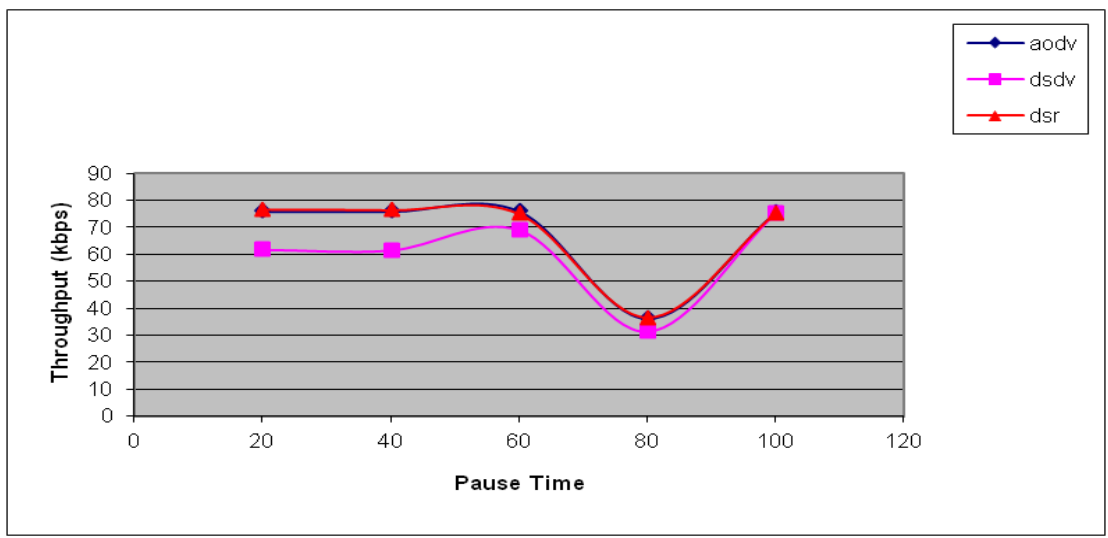

**Fig 7: Throughput under Pause Time (fixed 10 nodes).**

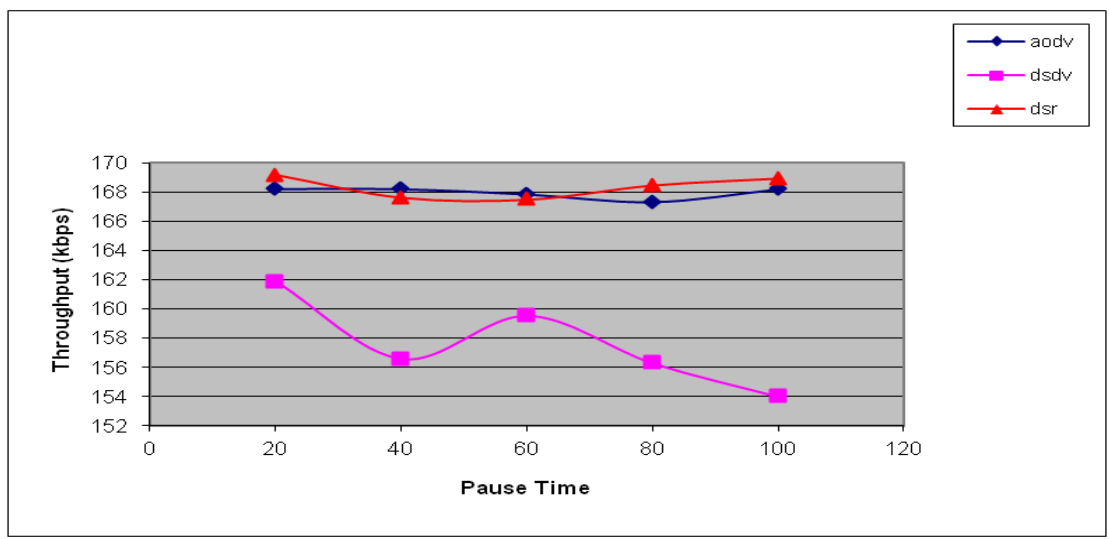

**Fig 8: Throughput under Pause Time (fixed 30 nodes).**

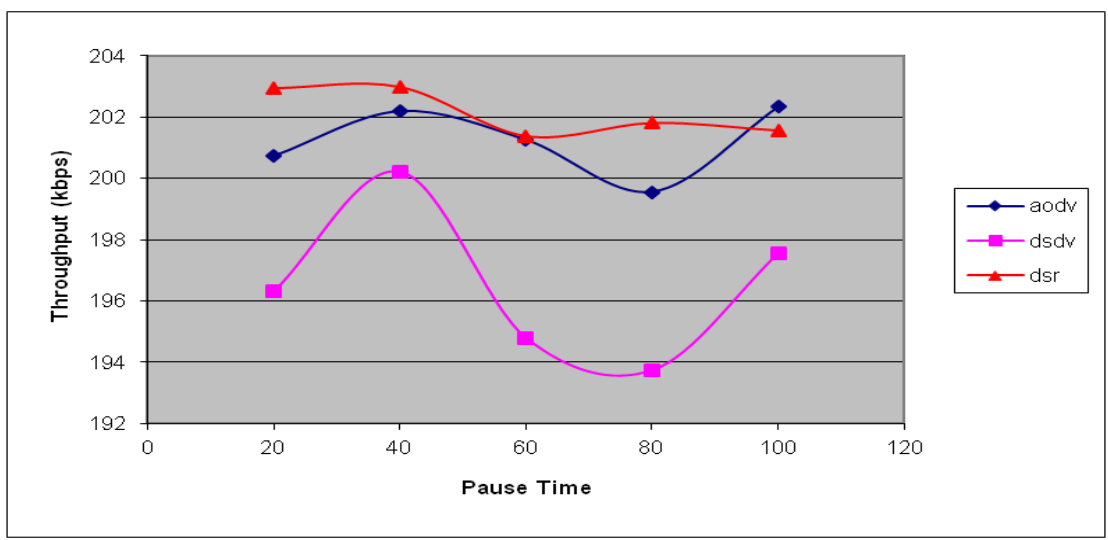

**Fig 9: Throughput under Pause Time (fixed 50 nodes).**

## 4.2 Simulation Analysis

In this paper, we have attempted to compare all the three protocols under the Manhattan Grid mobility model.

For all the simulations, the same movement models were used, the number of traffic sources was fixed at 10, 30 and 50, the pause time was varied as 0, 20, 40, 60, 80 and 100s, and a fixed topology boundary of 500x500.

As shown in figures 1, 2 and 3, we observe that, regardless of network size or mobility rate, AODV and DSR performed better than DSDV delivering over 90% of data packets.

The Average end-to-end Delay of packet delivery was higher in AODV as demonstrated in figures 4, 5 and 6.

Same figures show a uniform distribution of Average end-to-end Delay in DSDV and DSR which performed well than AODV.

Throughput was similar for both AODV and DSR and slightly higher as compared to DSDV (figures 7, 8, and 9).

Network size and network load have lead to increasing the throughput for the three protocols.

With increasing network size, we assume that under Manhattan Grid mobility model DSDV and DSR performs well than AODV by providing acceptable Average end to end Delay, throughput and packet delivery fraction (figure 3, 6 and 9).

## 5. CONCLUSION

In this paper, AODV, DSDV and DSR routing protocols using different parameter metrics have been simulated and analysed in terms of Packet Delivery Fraction (PDF), Average end-to-end Delay and Throughput in different environments.

Simulation results show that performance parameters of the routing protocols may vary depending on network load, mobility and network size.

Under Manhattan Grid mobility Model, AODV and DSR experience the highest Packet Delivery Fraction and Throughput with the increase of nodes pause time, CBR traffic sources and mobile nodes number. However, DSDV experiences the lowest Average end-to-end Delay.

AODV and DSR performance is due to their on demand characteristics to determine the freshness of the route. And it is proved also that AODV has a slightly higher Average end-to-end Delay than DSR.

In this paper, three routing protocols are used and their performances have been analyzed under Manhattan Grid mobility model with respect to three performance parameters. This paper can be enhanced by treating other MANET routing protocols under different mobility scenario with respect to other performance metrics.